%% file: main.tex
\documentclass[12pt]{article}
\usepackage{graphicx}
\usepackage{amsfonts}

\newcommand\pubdate{\today}

\newcommand\pubnumber{}

\def\Title#1{\begin{center} {\Large #1 } \end{center}}
\def\Author#1{\begin{center}{ \sc #1} \end{center}}
\def\Address#1{\begin{center}{ \it #1} \end{center}}

\newcommand\pubblock{\rightline{\begin{tabular}{l} \pubnumber\\
         \pubdate  \end{tabular}}}
\newenvironment{Abstract}{\begin{center}{\bf Abstract}\end{center} \bigskip \begin{quotation}  }{\end{quotation}}
\newenvironment{Presented}{\begin{quotation} \begin{center} 
             PRESENTED AT\end{center}\bigskip 
      \begin{center}\begin{large}}{\end{large}\end{center} \end{quotation}}
\def\Acknowledgements{\bigskip  \bigskip \begin{center} \begin{large}
             \bf ACKNOWLEDGEMENTS \end{large}\end{center}}
\input econfmacros.tex
\begin{document}
\begin{titlepage}
\pubblock

\vfill

%%%%%%%%%%%%%%%%%%%%%%%%%%%%%%%%%%%%%%%%%%%%%%%%%%%%%%%
%%MODIFY
%%%%%%% TITLE, AUTHOR, ADDRESS 
%%%%%%%%%%%%%%%%%%%%%%%%%%%%%%%%%%%%%%%%%%%%%%%%%%%%%%%

\Title{Top physics with the ATLAS detector}
\vfill
\Author{C. Helsens}
\centering \emph{On behalf of the ATLAS collaboration}
\Address{Institut de F\`isica d'Altes Energies (IFAE), Edifici Cn, Universitat Aut$\grave{o}$noma de Barcelona (UAB), E-08193 Bellaterra (Barcelona), Spain}
\vfill

%%%%%%%%%%%%%%%%%%%%%%%%%%%%%%%%%%%%%%%%%%%%%%%%%%%%%%%
%%MODIFY
%%%%%%% Abstract
%%%%%%%%%%%%%%%%%%%%%%%%%%%%%%%%%%%%%%%%%%%%%%%%%%%%%%%

\begin{Abstract}

During the 2010 CERN Large Hadron Collider operation at $\sqrt{s}$ = 7 TeV, 35 pb$^{-1}$ of high $p_T$ triggers has been collected
by the ATLAS detector. This corresponds to the production of approximately $\number 2500$ top-quark pair events containing at least one lepton 
in the final state ($e$ or $\mu$).

\end{Abstract}

\vfill

\begin{Presented}
The Ninth International Conference on\\
Flavor Physics and CP Violation\\
(FPCP 2011)\\
Maale Hachamisha, Israel,  May 23--27, 2011
\end{Presented}
\vfill

\end{titlepage}
\def\thefootnote{\fnsymbol{footnote}}
\setcounter{footnote}{0}
%

%%%%%%%%%%%%%%%%%%%%%%%%%%%%%%%%%%%%%%%%%%%%%%%%%%%%%%%
%%%%%%% Article body
%%%%%%%%%%%%%%%%%%%%%%%%%%%%%%%%%%%%%%%%%%%%%%%%%%%%%%%

\section{Introduction}

Top-quark precision measurements are of central importance to the LHC physics program. The top-quark the heaviest known
fundamental particle with unique properties well defined by the SM. It has large couplings to the Higgs boson, and it is the only quark that 
decays before hadronisation. The production of top-quark pairs in pp collisions is a process at the boundary between the
Standard Model (SM) and what might lie beyond it. Thus new physics might affect its properties. Within the SM, top quarks are predicted to 
almost always decay to a $W$-boson and a $b$-quark. The decay topologies are classified according to the $W$-boson decays. 
When top-quarks are pair produced, the branching ratios for the single-lepton and dilepton mode are 37.9$\%$ and 6.5$\%$ respectively. 
The corresponding final state signatures involve one or two leptons, (only electron or muon are considered), jets, 
two of them from b-quarks, and missing transverse energy ($E^{miss}_T$) from neutrinos from the W-boson decay

All the analyses presented in this note have been performed using 35 pb$^{-1}$ of data collected with the ATLAS detector~\cite{ref:ATLAS} 
in 2010. All of them have been cross-checked by independent analysis/methods and only the ones that give the best performance are 
presented here. Full documentation can be found in the notes given as references.

\section{Top-quark standard objects definition}
\label{sec:objID}
This section describes the general object definition used in top-quark analyses.
For the cosmic and pileup rejection at least five tracks associated to the primary vertex are required. Jets are reconstructed using 
topological clusters and the Anti-kT algorithm (with a distance parameter of R=0.4), calibrated at the particle level, with a 
$p_T >$ 25~GeV and pseudo-rapidity $\mid \eta \mid <$~2.5. Single lepton triggers are being used with a threshold of 15~GeV for 
electron and 13~GeV for muons. 
Electrons are defined using good isolated calorimeter objects, matched to a track in the inner detector with a transverse energy greater 
than 20~GeV, and $\mid \eta \mid<$ 2.47 removing the crack region between the barrel and end cap calorimeters [1.37,1.52].
Muons should also be combined objects with a segment in the tracker and the muon spectrometer; the track should be isolated with 
$p_T >$ 20~GeV and $\mid \eta \mid<$ 2.5.
When b-tagging is used to identify b-jets, a simple and robust (with a modest rejection factor) secondary vertex algorithm ~\cite{ref:SV0} 
is used with 
an average efficiency of 50$\%$ for $b$-jets. For most of the single lepton analyses a cut on the missing transverse energy and the 
transverse mass of the W are performed separately (electron channel $E^{miss}_T > $ 35~GeV and $m_T(W) >$ 25~GeV) or combined 
(muon channel $E^{miss}_T > $ 20~GeV and $E^{miss}_T + m_T(W) >$ 60~GeV).

\section{Top-quark pair production cross-section $\sigma_{t\bar{t}}$}
\label{sec:topXS}

\subsection{Single-lepton channel}
In the single-lepton channel two complementary measurements have been performed. For the first method, no explicit identification of
secondary vertices inside jets (b-tagging) is used~\cite{ref:SingleLeptonXS_Notag}. The main backgrounds are W+jets and QCD multi-jet 
events where one of the jets is misidentified as a lepton. The QCD multi-jet background is particularly difficult to simulate correctly and 
thus estimated using data-driven techniques. The event selection follows the standard top object definition (see Sect. ~\ref{sec:objID}). 
Events are classified according to the lepton flavor. Exactly one lepton of a given flavor is selected and the event is vetoed if a second 
lepton is found.
Two jet multiplicity bins are considered: 
exactly 3 jets and $\geq$ 4 jets. A likelihood discriminant is built following the projective likelihood approach based on three uncorrelated 
discriminating variables that exploit the different kinematic properties of the $t\bar{t}$ and the W+jets background. The variables are, 
the pseudo-rapidity of the lepton, the charge of the lepton and the aplanarity\footnote{exp(-8$\times \mathcal{A}$) with 
$\mathcal{A} = \frac{3}{2}\lambda_3$ where $\lambda_3$ is the smallest eigenvalue of the normalized momentum tensor calculated using
the selected jets and lepton in the event}. Individual likelihood 
functions are defined for each of the 4 channels and are multiplied together in a combined fit to extract a cross-section of 
$\sigma_{t\bar{t}} = 171 \pm 17 (stat.)^{+20}_{-17}(syst.)\pm 7 (lumi.)$pb. The main systematic uncertainties are the amount of initial and 
final state radiation, as well as the limited understanding of the jet energy scale and reconstruction efficiency. Cross-check 
measurements are performed with a one dimensional kinematic fit to the reconstructed top mass and cut-and-count methods. Both are found 
to be in good agreement with the main result.

A second method exploits b-tagging information in the context of a multivariate likelihood discriminant~\cite{ref:SingleLeptonXS_btag}. Four 
variables are used to construct the input multivariate likelihood template distributions, among which the average of the weights of the 
two most significant b-tags (jets with the lowest probability to originates from the primary vertex; tagger based on the signed 
impact parameter significance $d_0/\sigma_0$~\cite{ref:JetProb}). 
The other variables are the lepton pseudo-rapidity, the aplanarity and $H_{T,3p}$\footnote{transverse energy of 
all jets except the two leading ones, normalized to the sum of absolute values of all longitudinal momenta in the event.}
 More channels are used here with an additional jet multiplicity bin (3, 4 and $\geq$ 5 jets). 
A profile likelihood fit with 17 nuisance parameters combining the six channels is performed to extract 
the $\sigma_{t\bar{t}}$ and constrain the effect of systematic uncertainties using data. A cross-section of
$\sigma_{t\bar{t}} = 186 \pm 10 (stat.)^{+21}_{-20}(syst.)\pm 6 (lumi.)$pb is extracted, where the main systematic uncertainties originate  
from the b-tagging algorithm calibration from data and heavy flavor fraction in W+jets events. The result was cross-checked using a 
cut-and-count method and two one dimensional kinematic fits to the reconstructed top mass. Good agreement was also found.

\subsection{Dilepton channel}
In the dilepton channel the cross-section is extracted using a cut-and-count method~\cite{ref:DileptonXS} and no b-tagging requirement for 
the baseline analysis. Event candidates are selected by requiring exactly two opposite-signed leptons, among the three channels, 
$ee$, $\mu\mu$ and $e\mu$. The main background contribution is Drell-Yan production, and it is suppressed by requiring same-flavor events 
to have $E^{miss}_T > $ 40~GeV and imposing a Z-boson mass veto by excluding events with $\mid m_{ll} - m_Z \mid <$ 10~GeV; for $e\mu$ events, the 
scalar sum of jet and lepton transverse energies($H_T$) is requiered to be greated than 130~GeV. The remaining Drell-Yan and fake lepton 
contributions are estimated via data-driven techniques. The cross-section is extracted using a profile likelihood with a simultaneous fit 
to the three channels. A cross-section of $\sigma_{t\bar{t}} = 174 \pm 23 (stat.)^{+19}_{-17}(syst.)\pm 7(lumi.)$pb is obtained. 

In order to corroborate this measurement, four other independent methods have been developed. Two of then do not relly on the use of b-tagging.
One normalizes the $t\bar{t}$ yields to the Z-boson decays while the second method uses template in the ($E^{miss}_T, N_{jets}$) plane.
Another two methods relly on the b-tagging information: 
cut-and-count and a simultaneous measurement of the b-tagging efficiency and $\sigma_{t\bar{t}}$. All measurements are found to be in good 
agreement with each other.

\subsection{Combination}
\begin{figure}[htb]
\centering
\includegraphics[width=0.4\textwidth]{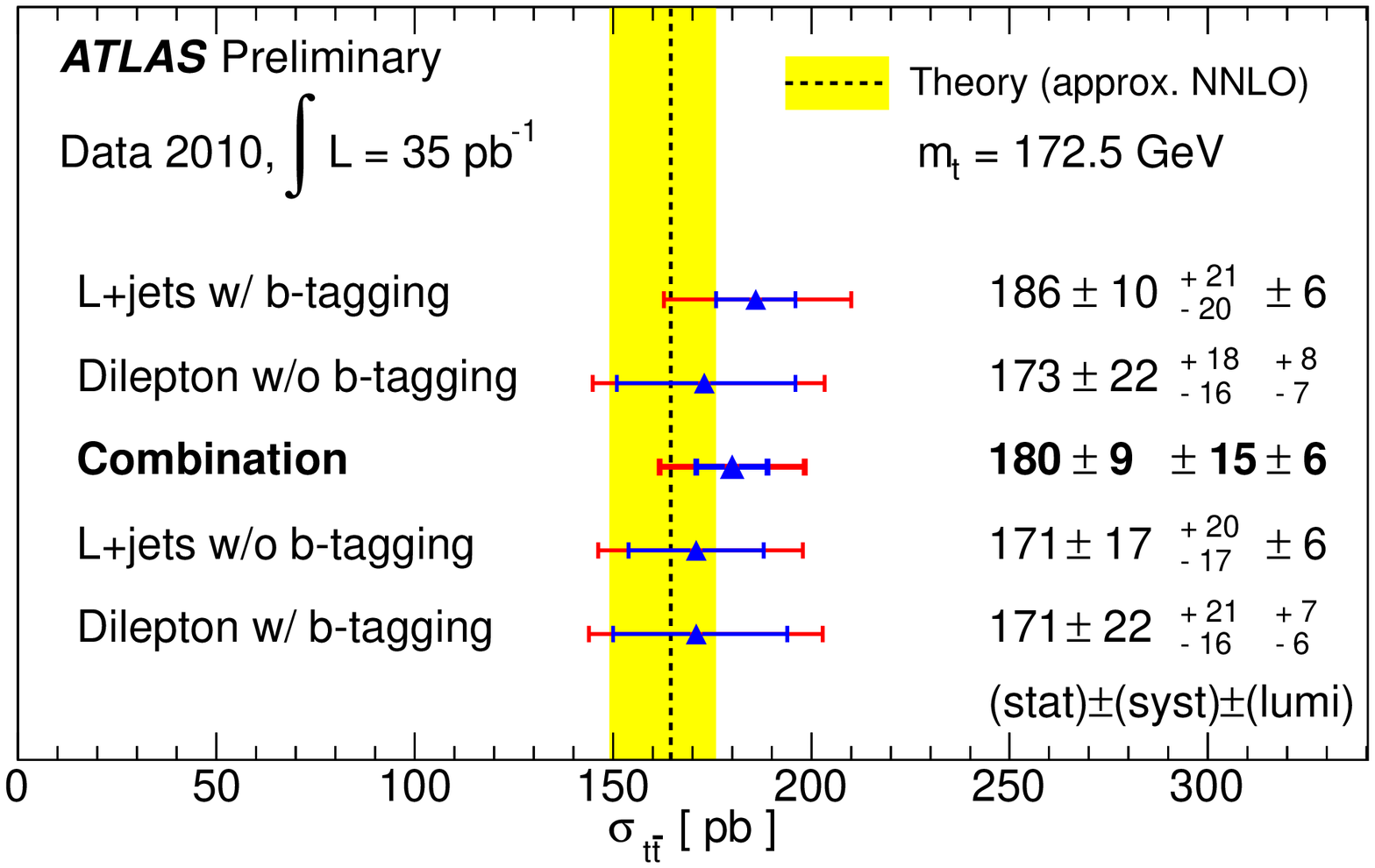} 
\includegraphics[width=0.4\textwidth]{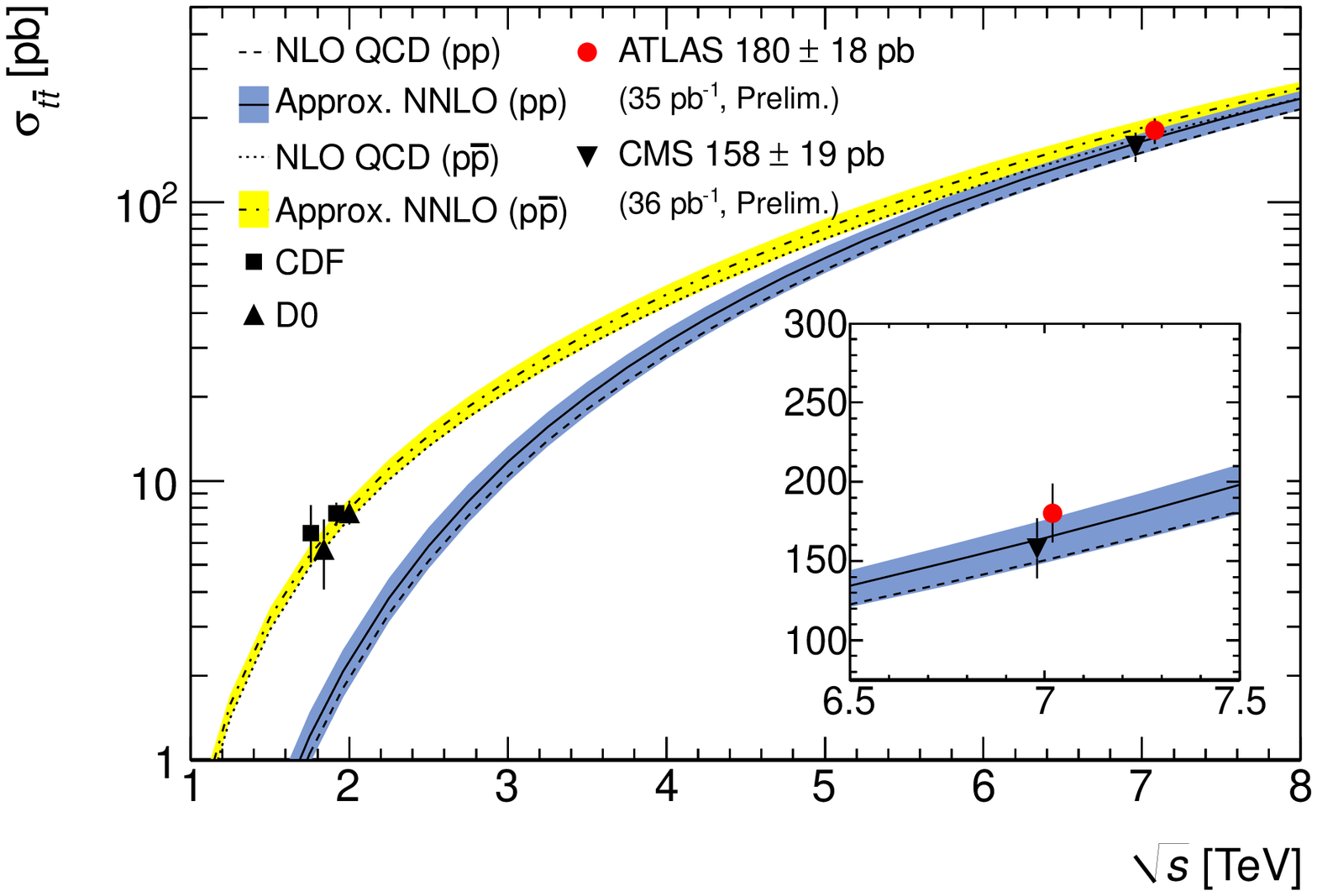} 
\caption{Comparison of the various cross-sections and combination compared to NNLO predictions (left).
Cross-section measurement at the Tevatron and LHC compared to approximate NNLO predictions (right).}
\label{fig:XScomb}
\end{figure}

The two most precise cross-section measurements in the single-lepton and dilepton channels are combined~\cite{ref:CombinationXS}: 
single-lepton with b-tagging and dilepton cut-and-count without b-tagging. The combined result, taking into 
account correlations in the systematic uncertainties gives a measurement with a total uncertainty of 10$\%$, 
$\sigma_{t\bar{t}} = 180 \pm 9(stat.) \pm 15(syst.)\pm 6(lumi.)$pb, and is in excellent agreement with the standard model prediction 
as shown in Fig.~\ref{fig:XScomb}.

\subsection{Fully hadronic}
The $\sigma_{t\bar{t}}$ production cross-section in the fully hadronic channel characterized by a six jets topology in the final state
is a challenging measurement. This channel has the advantage of a large branching ratio, 46$\%$, although is suffers from a large QCD 
multi-jet background~\cite{ref:FullhadXS}. A cross-section of $\sigma_{t\bar{t}} = 118 \pm 73(stat.) \pm 48(syst.)\pm 4(lumi.)$pb is 
measured. The result is not yet competitive but it represents the first top-quark measurement in this channel at the LHC.

\section{Single-top production}
The first observation of the single-top quark production was repported in 2009 by both the D0~\cite{ref:stopD0} 
and CDF~\cite{ref:stopCDF} experiments. This final state provides a direct probe of the $W-tb$ coupling and is thus sensitive to many models
of new physics~\cite{ref:2newphysStop}. The measurement of the production cross-section determines the magnitude of the quark mixing matrix 
element $V_{tb}$ without assumptions about the number of quark generations. 

In the $t-$channel the search is based on the selection of events with a single lepton, jets and $E^{miss}_T$~\cite{ref:sinleTop}. 
This is an analysis based on the application of sequential cuts. Among a large number of kinematic variables considered, it was 
found that imposing requirements on the pseudo-rapidity of the leading untagged jet and the reconstructed top-mass, a high significance, including 
statistical and systematics uncertainties on background estimates can be reached. This selects a total of 32 events. The QCD multi-jet 
and W+jets backgrounds are estimated using a data-driven method, and a cross-section of 
$\sigma_{t} = 53^{+27}_{-24}(stat.)^{+38}_{-27}(syst.)$pb, with a significance of 1.6$\sigma$ is estimated. 

The $Wt$-channel associated production is more difficult to observe than the $t-$channel production due to its smaller 
cross-section~\cite{ref:sinleTop}. It contains two W-bosons in the final state. The single and dilepton channels are combined and 
a 95$\%$ confidence level upper limit is set on the $Wt$-channel production cross-section of $\sigma_{Wt} < 158$~pb.

\section{Top-quark properties}
\subsection{Mass}
The top-quark mass is a fundamental parameter of the SM. It enters in radiative corrections to precision electroweak observables 
used to derive constraints on the as yet undiscovered Higgs boson. The current world average is 
$m_{top} = 173.3 \pm 1.1$~GeV~\cite{ref:bestTopMass}. Two methods are used in ATLAS to extract the top-quark mass: 
fitting templates to the reconstructed mass and deriving the mass from the measured $\sigma_{t\bar{t}}$.

\subsubsection{From direct reconstruction}
The top-quark mass can be extracted from template fits to observables derived from kinematics of the reconstructed final state. The key 
point for the template method is the precise knowledge of the uncertainty in the  jet energy scale (JES). Three complementary template 
analysis have been developed that address the JES systematic in different ways~\cite{ref:topmasstemplate}. The baseline method is a 
1-dimensional analysis based on the measurement of the mass ratio $R_{32} = m_{jjj}/m_{jj}$  from hadronic top-quark candidates 
(approximate the ratio of the top-quark mass to the W-boson mass). Due to the stability of $R_{32}$ against JES variations, 
this method avoids the need to use in-situ calibrations techniques to reduce the impact of JES. Templates are derived for 
the $R_{32}$ distribution for signal and backgrounds. A combined fit to the two lepton flavor decay channels provides a 
top-quark mass measurement of $m_{top} = 169.3 \pm 4.0 (stat.)\pm 4.9 (syst.)$~GeV. A different method consists in simultaneously 
extracting $m_{top}$ and a global jet energy scale factor between data and predictions. And a last method uses a 1-dimensional 
kinematic fitter to all the decay products of the $t\bar{t}$ system.

\subsubsection{From cross-section measurement}
The direct measurement of the top-quark mass from templates depends on the Monte-Carlo (MC) simulation either to fit the chosen kinematic 
observable~\cite{ref:topmasstemplate} or to calibrate the measurement~\cite{ref:massD0,ref:massCDF}. In the MC, the top-quark mass does 
not correspond to a well defined renormalization scheme leading to an uncertainty in its definition.

Another possibility to measure the top-quark mass is to derive it from the $t\bar{t}$ cross-section. This analysis compares the measured 
inclusive $\sigma_{t\bar{t}}$ in the lepton + jet channels~\cite{ref:SingleLeptonXS_btag} with fully inclusive higher order perturbation 
QCD computation where the top-quark mass is defined as the pole mass. The extraction of $m_{pole}^{top}$ from cross-section measurement 
provides complementary information compared to direct measurement methods that rely explicitly on the detailed kinematic reconstruction. 
This extraction also tests the internal consistency of perturbative QCD calculations for $\sigma_{t\bar{t}}(m_{pole}^{top})$ in 
a well defined renormalization scheme. The theoretical prediction for the cross-section used in this extraction are calculated in various 
approaches using the pole mass definition for the top-quark, namely in approximate NNLO (NNLO Langenfeld~\cite{ref:Langenfeld}, 
NNLO-Kidonakis~\cite{ref:Kidonakis}, or NLO + NNLL (NNLL-Ahrens~\cite{ref:Ahrens})). 
For this extraction, these were parametrized as:
\begin{equation}
 \label{eq:massXS}
 \sigma_{t\bar{t}}(m_{pole}^{top}) = \Bigg( \frac{1}{m_{pole}^{top}} \Bigg)^4 \Big(a + b(m_{pole}^{top} - 170) + c(m_{pole}^{top} - 170)^2 + d(m_{pole}^{top})^3 \Big)pb,
\end{equation}
To extract the top-quark mass a combined uncorrelated theoretical/experimental likelihood is constructed using Gaussian probability 
density function. The maximum of this likelihood determines the extracted $m_{pole}^{top}$. The main result is obtained using the 
NNLO-Langenfeld~\cite{ref:Langenfeld} prediction and is found to be $m_{pole}^{top} = 166.4^{+7.8}_{-7.3}$~GeV.

\subsection{W-boson helicity in top-quark decays}
The polarization states of the W-boson in top-quark decays are well defined in the SM, due to the $V - A$ structure of the charged current 
weak interactions. The SM predicts that $\sim70\% (\sim30\%)$ of $W$-bosons in top-quark decays are longitudinal (left-handled). 
In this analysis~\cite{ref:Whel} the W-boson helicity fractions and angular asymmetries are measured in top-quark pair production 
in single-lepton channel exploiting the angular distributions of the $t \rightarrow bW \rightarrow bl\nu_l$ decay products. 
The distribution of $\theta^*$ (angle between the direction of the lepton and the reversed momentum of the b quark from the 
top-quark decay, both boosted into the W-boson rest frame), is distorted by the detector response, kinematic cuts, off-line event 
selection and reconstruction. 

In a first measurement templates of $cos\theta^*$ are created and used to fit the data to extract the W-boson helicity fractions. Assuming 
$F_R = 0$, helicity fractions are extracted from the data and found to be $F_L = 0.41 \pm 0.12$ and $F_0 = 0.59 \pm 0.12$ respectively.

The second measurement is based on angular asymmetries constructed from the $cos\theta^*$  variable. Selected event are reconstructed 
using a $\chi^2$ fit and an iterative procedure is applied to correct for detector and reconstruction effect in order to recover the 
undistorted distribution at the parton level. Helicity fractions are measured to be $F_L = 0.36 \pm 0.10$, $F_0 = 0.65 \pm 0.15$ 
and $F_R = -0.01 \pm 0.07$. Both results are in very good agreement with the SM predictions and are used to place limits on anomalous 
couplings $V_R, g_L$ and $g_R$ that arise in beyond the SM models of new physics.

\subsection{Search for anomalous $t\bar{t} + E^{miss}_{T}$ events}
A search for anomalous $t\bar{t} + E^{miss}_T$ in the single-lepton final state has been performed~\cite{ref:highMET}. Such phenomena can 
arise from a number of SM extensions, but we focused here on the search for a pair produced top partner decaying to a top-quark and a 
long lived neutral particle which escape undetected. The benchmark model considered is top-quark partner production $T$ with 
$T \rightarrow tA_0$, $A_0$ being a Dark Matter (DM) candidate. The final state is identical to the SM $t\bar{t}$ case but with 
a larger $E^{miss}_{T}$. Comparing the data and the model, a good agreement between the SM and the data is found (17 event observed, 17.2 expected), and exclusion 
is performed, $m(T) < 300$~GeV for $m(A_0) = 10$~ GeV and $m(T) < 275$~GeV for $m(A_0) = 50$~ GeV at 95$\%$ C.L.

\subsection{Search for high mass $t\bar{t}$ events}

A search for high mass phenomena producing top-quark pairs has been performed~\cite{ref:ttbarres}. Two benchmark models are considered: 
the search for heavy particle decaying to $t\bar{t}$ pairs (leptophobic $Z'$)~\cite{ref:leptophobic} and 
a search for quantum black holes (QBH)~\cite{ref:QBH} through $t+X$ at high mass. Both analysis are using the single lepton channel. No evidence of a resonance 
has been found so far and we excluded $m_{QBH} < $2.35~TeV at 95$\%$ C.L. and an observed 95$\%$ C.L. cross-section upper limit of 55(2.2)pb 
for $m_{Z'} = 500(2000)$~GeV

\section{Conclusion}
With the first 35~pb$^{-1}$ of proton collisions that have been collected at $\sqrt{s} = 7$~TeV in 2010 by ATLAS, a suite of measurements 
involving top-quarks have been performed. The precision of the $t\bar{t}$ cross-section measurement is already competitive with the accuracy
of the theoretical predictions at the 10$\%$ level. An increase of integrated luminosity by two orders of magnitude expected by the end of 
2011 will allow improved precision in all the statistics limited analyses and the whole spectrum of top physics, including new physics searches 
will be explored at the LHC.

\Acknowledgements
I whish to thanks all the LHC teams and the ATLAS Collaboration for the wonderfull performances of the accelerator complex and the 
excellent data quality, the many authors and contributors of top physics analyses shown here, Martine Bosman and Aurelio Juste for 
reading and commenting this manuscript and the organizers of the ninth edition of \emph{Flavor Physics and CP Violation} conference. 
I also gratefully acknowledge the support of my institute for the fundings.

\end{document}

%% file: econfmacros.tex
\def\beq{\begin{equation}}
\def\eeq#1{\label{#1}\end{equation}}
\def\eeqn{\end{equation}}

\def\beqa{\begin{eqnarray}}
\def\eeqa#1{\label{#1}\end{eqnarray}}
\def\eeqan{\end{eqnarray}}

\let\bar=\overbar

\def\Dslash{\not{\hbox{\kern-4pt $D$}}}
\def\dslash{\not{\hbox{\kern-2pt $\del$}}}

\def\msb{{\bar{\ssstyle M \kern -1pt S}}}